\documentclass[a4paper,11pt]{article}
\pdfoutput=1

\usepackage{amsmath}
\usepackage{amssymb}
\usepackage[hidelinks]{hyperref}
\hypersetup{
    bookmarks=black,%
    colorlinks,%
    citecolor=blue,%
    filecolor=black,%
    linkcolor=black,%
    urlcolor=blue
}
\usepackage[labelfont=bf]{caption}
\usepackage{subcaption}
\usepackage{cite}
\usepackage{tikz}
\usepackage{ifthen}
\usepackage{bm}
\usepackage{authblk}
\usepackage{varwidth}
\usepackage{color}
\usepackage{braket}
\usepackage[utf8]{inputenc}

\newcommand{\beq}{\begin{equation}}
\newcommand{\eeq}{\end{equation}}
\newcommand{\bal}{\begin{aligned}}
\newcommand{\eal}{\end{aligned}}
\newcommand{\rmd}{\mathrm{d}}

\numberwithin{equation}{section}

\title{\bf The Price of Curiosity: \\Information Recovery in de Sitter Space
}

\author[a]{Lars Aalsma\thanks{\href{mailto:laalsma@wisc.edu}{\texttt{laalsma@wisc.edu}}}}
\author[b]{Watse Sybesma\thanks{\href{mailto:watse@hi.is}{\texttt{watse@hi.is}}}}

\affil[a]{\protect\begin{varwidth}[t]{\linewidth}\protect\centering Department of Physics, University of Wisconsin-Madison, \par 1150 University Ave, Madison, WI 53706, USA\protect\end{varwidth}}

\affil[b]{\protect\begin{varwidth}[t]{\linewidth}\protect\centering 
Science Institute, University of Iceland, \par Dunhaga 3, 107 Reykjav\'{i}k, Iceland
\protect\end{varwidth}}

\date{}

\begin{document}

\maketitle
\vspace{2em}
\begin{abstract}
    \noindent
Recent works have revealed that quantum extremal islands can contribute to the fine-grained entropy of black hole radiation reproducing the unitary Page curve. In this paper, we use these results to assess if an observer in de Sitter space can decode information hidden behind their cosmological horizon. By computing the fine-grained entropy of the Gibbons-Hawking radiation in a region where gravity is weak we find that this is possible, but the observer's curiosity comes at a price. At the same time the island appears, which happens much earlier than the Page time, a singularity forms which the observer will eventually hit. We arrive at this conclusion by studying Jackiw-Teitelboim gravity in de Sitter space. We emphasize the role of the observer collecting radiation, breaking the thermal equilibrium studied so far in the literature. By analytically solving for the backreacted geometry we show how an island appears in this out-of-equilibrium state.
\end{abstract}
\vfill

\newpage
\tableofcontents

\section{Introduction}
Early studies of thermodynamics in de Sitter space revealed that the cosmological horizon surrounding a static observer has thermodynamic properties very similar to those of a black hole \cite{Gibbons:1977mu}. The cosmological horizon has a temperature proportional to the surface gravity, an entropy proportional to its area and also obeys similar thermodynamic relations. In contrast, the microscopic interpretation of these properties are less clear. In particular, it is has been debated if and how a finite-dimensional Hilbert space can be implemented in de Sitter space, see e.g \cite{Witten:2001kn,Goheer:2002vf,Parikh:2004wh,Parikh:2004ux,Parikh:2002py,Anninos:2011af,Anninos:2017eib,Dong:2018cuv}.

Nonetheless, in spite of the absence of a fully understood microscopic description of de Sitter space we can still make progress just using the semi-classical approximation. In black hole physics, it has long been appreciated that the finiteness of black hole entropy places constraints on the process of black hole evaporation even in the semi-classical regime. Famously, following Hawking's computation the (entanglement) entropy of an evaporating black hole naively seems to grow larger than the decreasing Bekenstein-Hawking entropy of the black hole. This conflicts the reasonable assumption that black hole evaporation can be a unitary process in which an initially pure state with vanishing entanglement entropy cannot evolve towards a mixed state with non-zero entropy. Instead, Page showed that unitarity implies that the entanglement entropy of the radiation initially grows but decreases when roughly half of the black hole has evaporated \cite{Page:1993wv,Page:2013dx}. This apparent tension between unitarity and the thermal nature of Hawking radiation is known as the information paradox.

Only recently, it has been found how this paradox might be resolved within semi-classical gravity. In a series of breakthroughs \cite{Ryu:2006bv,Hubeny:2007xt,Faulkner:2013ana,Barrella:2013wja,Engelhardt:2014gca,Penington:2019npb,Almheiri:2019psf} following the increasing understanding of entanglement entropy in holography, it has been found that the fine-grained entropy of a non-gravitating region entangled with a gravitational system can receive contributions from a disjoint (island) region that lies in the gravitational system.\footnote{Subtleties with respect to how the non-gravitational region is coupled (as an external bath or as part of the spacetime) have been pointed out in \cite{Laddha:2020kvp}. In \cite{Geng:2020fxl} the situation was considered where the ``reservoir'' in which radiation is collected is gravitating.} When computing the fine-grained entropy using the replica trick, the contribution of the island appears as a new saddle point to the Euclidean gravitational path integral that was previously not taken into account. When the entanglement between the non-gravitating and gravitating region becomes large, this new saddle point gives the dominant contribution (giving the lowest entropy). By applying the island formula to the Hawking radiation of an evaporating black hole, the fine-grained entropy is found to follow the Page curve, initially increasing in accord with Hawking's computation and later decreasing when the new saddle point dominates. These results provide compelling evidence that black hole evaporation is a unitary process, which ensures that any information thrown into a black hole can eventually be recovered. By now, the island formula has been applied to many different black holes in different setups \cite{Almheiri:2019hni,Almheiri:2019yqk,Almheiri:2019psy,Gautason:2020tmk,Anegawa:2020ezn,Hashimoto:2020cas,Hartman:2020swn,Hollowood:2020cou,Dong:2020uxp,Chen:2020tes,Hartman:2020khs,Balasubramanian:2020hfs,Balasubramanian:2020xqf,Balasubramanian:2020coy,Alishahiha:2020qza,Chen:2020jvn,Geng:2020qvw,Chen:2020uac,Chen:2020hmv,Harlow:2020bee,Hernandez:2020nem,Akal:2020twv,Basak:2020aaa,Caceres:2020jcn,Karananas:2020fwx,Wang:2021woy} and these developments have been reviewed in \cite{Almheiri:2020cfm,Marolf:2020rpm,Raju:2020smc}.

Given the remarkable success of semi-classical physics in resolving the information paradox, one might wonder whether these developments can also shed light on the nature of cosmological horizons. In particular, we would like to understand if islands also contribute to fine-grained entropy in de Sitter space allowing a static observer to decode information thrown through the horizon. An additional complication is that in de Sitter space there is no asymptotic non-gravitating region where an observer can collect radiation in a subsystem to which the island formula can be applied. To evade this issue one can consider a set-up in which de Sitter space is entangled with a disjoint and non-gravitating universe or define a region at future infinity where the island formula can be applied \cite{Chen:2020tes,Hartman:2020khs,Balasubramanian:2020xqf}. The latter approach seems especially suitable to study inflationary physics. However, it is unclear if these approaches shed light on the experience of a static observer.

To explore if a static observer can decode information by performing measurements on the collected Hawking radiation, we therefore focus on the situation where they collect radiation inside of their static patch which breaks the thermal equilibrium studied so far when searching for islands \cite{Chen:2020tes,Hartman:2020khs,Balasubramanian:2020xqf,Sybesma:2020fxg,Geng:2021wcq}.\footnote{In particular, in \cite{Sybesma:2020fxg} the entanglement entropy of radiation in the entire static patch was considered without requiring radiation to be collected in a compact system.} To have analytical control over the backreacted geometry, we study this problem in the context of two-dimensional Jackiw-Teitelboim (JT) gravity on a de Sitter background. This has the additional advantage that analytical expressions for the entanglement entropy are known. Our description of an observer that collects radiation crucially relies on a non-equilibrium state first studied in \cite{Aalsma:2019rpt} that contains only incoming radiation, but no outgoing radiation leading to a pileup of positive energy. We analytically solve for the backreacted geometry and identify a region where gravity is weak, allowing us to use the island formula to compute the fine-grained entropy of the collected radiation. The entropy follows a Page curve, showing  that information behind the de Sitter horizon can in principle be recovered. However, precisely when the island first appears to a static observer (at a time parametrically smaller than the Page time) backreaction already leads to the formation of a singularity outside of their static patch. The observer's curiosity therefore comes at a price and after the island appears their fate is sealed. Their worldline eventually reaches a singularity where it terminates.

The rest of this paper is organized as follows. In Section \ref{sec:JTgravity}, we first review some well-known facts about JT gravity on a de Sitter background. We consider two different two-dimensional models that arise as distinct reductions from a higher-dimensional theory. After that, we construct the non-equilibrium state of interest and study the backreacted geometry. In Section \ref{sec:entropy} we compute the thermodynamic entropy of the radiation and cosmological horizon and use the island formula to compute the fine-grained entropy. After that, we use a quantum singularity theorem in Section \ref{sec:fate} to prove that a singularity forms in this state and discuss our results in Section \ref{sec:discussion}.

\section{JT gravity in de Sitter space} \label{sec:JTgravity}
Jackiw-Teitelboim (JT) gravity is a two-dimensional dilaton theory of gravity that can be obtained as a dimensional reduction of a higher-dimensional theory. In this section, we review some known aspects of JT gravity on a de Sitter background and consider two different models that have a distinct higher-dimensional origin. By coupling to a two-dimensional conformal matter sector we construct solutions in different vacuum states and study their corresponding backreacted geometry in detail.

\subsection{Vacuum solutions and coordinate systems}
The action of JT gravity coupled to a matter CFT on a de Sitter background with length scale $\ell$ is given by \cite{JACKIW1985343,TEITELBOIM198341}
\beq \label{eq:JTaction}
\bal
I &= \frac{\Phi_0}{2\pi}\left(\int \rmd^2x\sqrt{-g}R - 2\int\rmd x\sqrt{|h|} K\right) \\
&+\frac1{2\pi}\left(\int \rmd^2x\sqrt{-g}\,\Phi\left(R-\frac2{\ell^2}\right) - 2\int\rmd x\sqrt{|h|}\,\Phi K \right) + I_{\rm CFT}~.
\eal
\eeq
The first line multiplied by the constant $\Phi_0$ is a topological term and the second line contains the dynamical part of the dilaton $\Phi$. We will denote their sum by $\Phi_H = \Phi_0 + \Phi$. The equations of motion for the dilaton and the metric respectively are given by
\beq
\bal \label{eq:EOM}
R-2/\ell^2 &=0 ~, \\
\Phi g_{ab} - \ell^2 \nabla_a\nabla_b\Phi + \ell^2 g_{ab}\square \Phi &= \pi \ell^2\braket{T_{ab}} ~.
\eal
\eeq
$\langle T_{ab}\rangle$ is the quantum-corrected stress tensor of conformal matter coupled to the JT action. In the large $c$ limit, the one-loop corrections are dominated by the conformal anomaly.

\begin{figure}[ht]
\centering
\includegraphics[scale=.75]{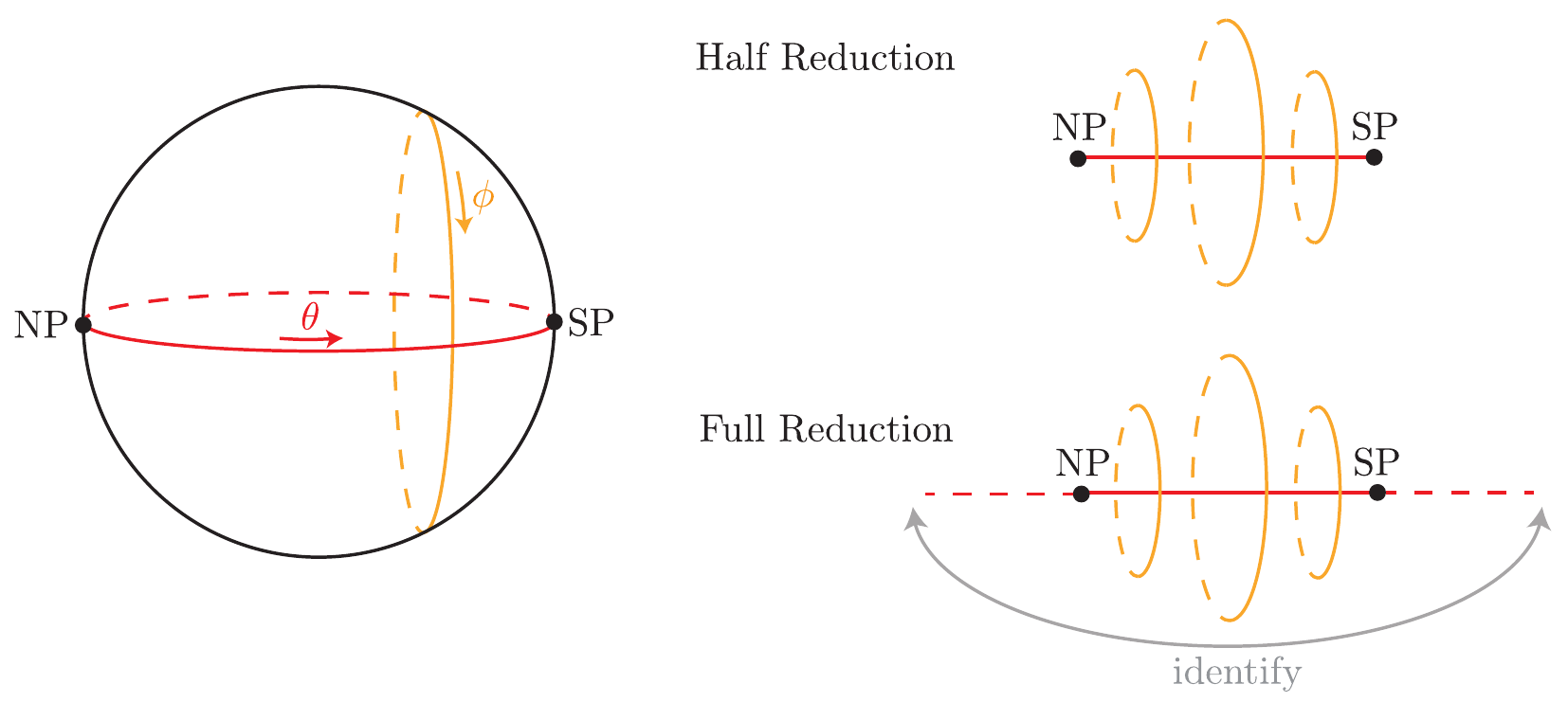}
\caption{\emph{Left:} Constant time slice of de Sitter space in static coordinates. The polar angle $\theta$ runs from $-\pi/2$ to $+\pi/2$ and the azimuthal angle $\phi=\phi+ 2\pi$. \emph{Right:} To obtain a two-dimensional dilaton gravity model we perform a reduction over the $S^1$ parametrized by $\phi$. We consider both a ``half reduction'' where we restrict $\theta$ to run from $-\pi/2$ to $+\pi/2$ and a ``full reduction'' where $\theta=\theta+2\pi$. The circumference of the orange $S^1$ indicates the size of the dilaton, which becomes negative in the red dotted region in the full reduction.}
\label{fig:Reduction}
\end{figure}

The JT action can be obtained by performing a reduction of a higher-dimensional theory. Explicitly, we will consider two distinct models. First, we start with three-dimensional de Sitter space in static coordinates
\beq
\rmd s^2=\left(1-r^2/\ell^2\right)\rmd t^2 + \left(1-r^2/\ell^2\right)^{-1}\rmd r^2 + r^2\rmd\phi^2 ~.
\eeq
We can perform a Kaluza-Klein reduction over the $S^1$ parametrized by $\phi$ (with the Kaluza-Klein photon set to zero) to obtain the JT action with $\Phi_0=0$ \cite{Sybesma:2020fxg}. The dilaton is then identified as the radial direction in the three-dimensional theory:
\beq
\Phi = \frac{\pi}{4G_3}r ~.
\eeq
It then follows that the vacuum equations of motion $(T_{ab}=0)$ are solved by
\beq \label{eq:staticdil}
\Phi = \frac{\Phi_s}{24} \cos\theta ~,
\eeq
where we introduced the coordinate $r=\ell\cos\theta$ in which the two-dimensional static line element becomes
\beq \label{eq:staticmetric}
\rmd s^2 = -\sin^2\theta\rmd t^2 + \ell^2\rmd \theta^2 ~.
\eeq
The constant $\Phi_s$ is defined with a factor of $1/24$ for later convenience. The three-dimensional de Sitter entropy is given by the dilaton evaluated on the horizon.
\beq
S_{\rm 3d} = \left.2\Phi_H\right|_{\theta=0} = \frac{\Phi_{s}}{12}~.
\eeq
In the three-dimensional origin of this model $\cos\theta\geq 0$ and imposing this on the two-dimensional theory we obtain a model in which $\Phi\geq0$. Doing so, spatial sections of the Penrose diagram only describe a semicircle which requires us to specify boundary conditions at the poles. We can equivalently think of this model as a $\mathbb{Z}_2$ orbifold of two-dimensional de Sitter space. We refer to this model as a ``half reduction'' in Figure \ref{fig:Reduction}. A similar partial reduction to JT gravity from a three-dimensional AdS geometry has been considered in \cite{Achucarro:1993fd} and more recently in \cite{Verheijden:2021yrb}.

\begin{figure}[ht!]
\centering
\includegraphics[scale=1]{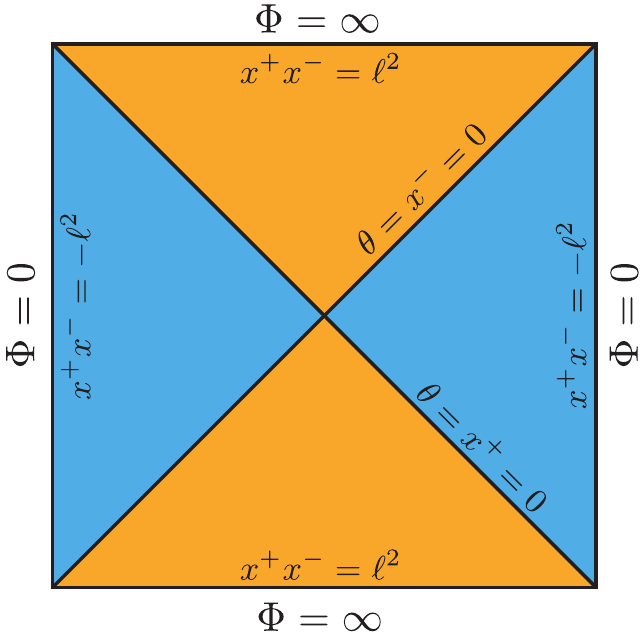}
\caption{Penrose diagram of two-dimensional de Sitter space obtained as a half reduction where we restrict $\Phi\geq0$ and impose reflecting boundary conditions at $\Phi=0$. The static coordinates \eqref{eq:staticmetric} cover the two static patches shaded in blue and the coordinates $x^\pm$ (see \eqref{eq:KruskalMetric}) cover both the blue and orange shaded regions.}
\label{fig:PenroseSquare}
\end{figure}

\begin{figure}[ht!]
\centering
\includegraphics[scale=1]{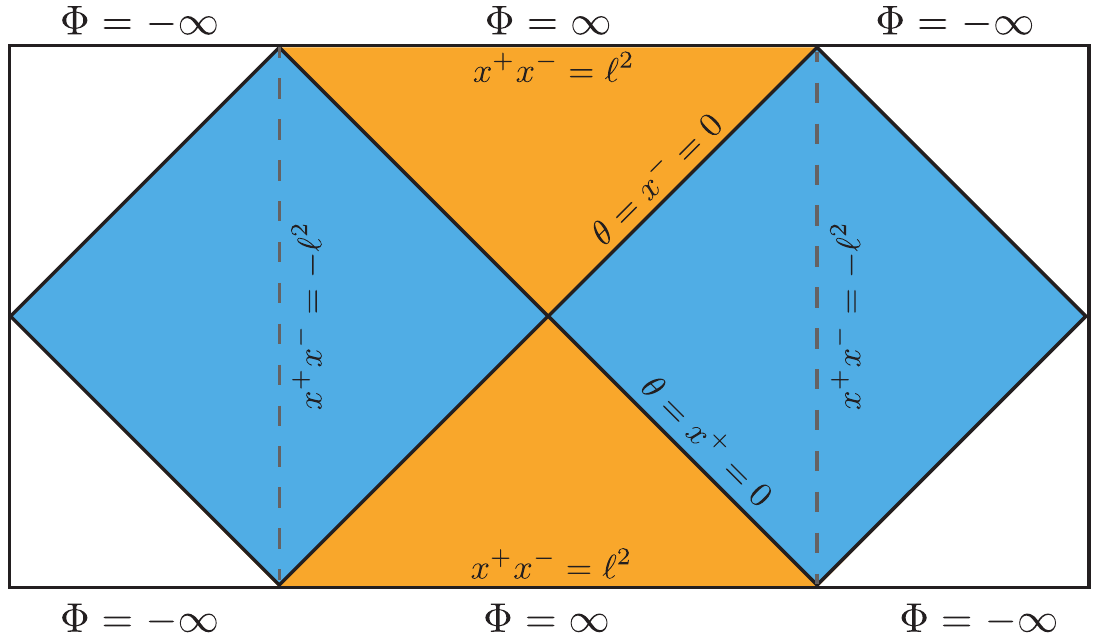}
\caption{Penrose diagram of two-dimensional de Sitter space obtained as a full reduction in the near-horizon limit of a Nariai black hole in four dimensions. The static coordinates \eqref{eq:staticmetric} cover the two static patches shaded in blue and the coordinates $x^\pm$ (see \eqref{eq:KruskalMetric}) cover both the blue and orange shaded regions. The white regions are the black hole interior and the places where $\Phi\to-\infty$ are its singularities. The left and right vertical lines of the diagram are identified.}
\label{fig:PenroseRectangle}
\end{figure}

Second, we also consider a ``full reduction'' (see Figure \ref{fig:Reduction}) where we do not impose the dilaton to take positive values. This model is more naturally obtained when we consider a spherical compactification of a Nariai black hole in four-dimensional de Sitter space, whose near-horizon geometry is dS$_2\times S^2$, close to the situation in which the black hole and cosmological horizon coincide. The combination $\Phi_H=\Phi_0+\Phi$ measures the area of the transverse sphere and the regions where $\Phi\to-\infty$ are interpreted as the black hole singularity. Because the geometry now contains both a black hole and cosmological horizon, the entropy in a single static patch is given by the sum of the horizons located at $\theta=(0,\pi)$. 
Adding both contributions the dynamical part of the dilaton vanishes and the entropy as seen from one static patch is given just by the topological part of the action. In Euclidean signature we have
\beq
S_{\rm 4d} =\frac{\Phi_0}{2\pi}\left(\int \rmd^2x\sqrt{g}R - 2\int\rmd x\sqrt{|h|} K\right) = 4\Phi_0 ~.
\eeq
Because the sum of the cosmological and black hole entropy of the Nariai geometry is given by $2/3$ of the entropy of empty four-dimensional de Sitter space, we can identify $2\Phi_0$ as $1/3$ of the entropy of four-dimensional de Sitter space. It is important to keep in mind that while $\Phi_0$ is proportional to the entropy, $\Phi$ measures the deviation away from the exact Nariai limit. The two-dimensional model is only valid close to this limit and for that reason $\Phi_{0}\gg \Phi_{s}$ and $\Phi_{0}\gg1$. The Penrose diagram of the half and full reduction are given in Figure \ref{fig:PenroseSquare} and \ref{fig:PenroseRectangle} respectively.

If we now include conformal matter, we will need to specify a quantum state to evaluate the vacuum expectation value of the stress tensor and to obtain the corrected solution to the equations of motion. In two dimensions, different vacuum states of a conformal matter sector and their respective stress tensors can be easily constructed in the following way. First, consider a set of null coordinates $x^\pm$ and expand the field operator in a complete set of left and right-moving modes that are positive frequency with respect to these coordinates. Expressed in terms of these coordinates, the diagonal components of the stress tensor vanish (because we are considering a vacuum state) and the off-diagonal component is fixed by the state-independent conformal anomaly. On a de Sitter background this takes the form \cite{Candelas:1978gf}
\beq
\braket{T^a_{\,\,\,a}} = \frac{c}{12\pi\ell^2} ~.
\eeq
The conformal anomaly can be obtained explicitly by computing the one-loop effective action which gives rise to a Polyakov term \cite{Callan:1992rs}. Here $c$ is the central charge of the CFT. By taking $c\gg 1$ loop corrections involving e.g. the dilaton are suppressed compared to those involving the conformal matter\cite{Callan:1992rs}. However, for the latter effects to not dominate over the classical contribution we have to take $\Phi_s\gg c$. We stress that the CFT is intrinsically two-dimensional as in that it does not have a (known) higher-dimensional counterpart.

After having fixed a vacuum state, we can express the stress tensor in a different set of null coordinates $y^\pm(x^\pm)$ using the anomalous transformation law
\beq \label{eq:stresstrafo}
\langle T_{\pm\pm}(y^\pm)\rangle = ({x^\pm}')^2\langle T_{\pm\pm}(x^\pm)\rangle -\frac{c}{24\pi}\{x^\pm,y^\pm\} ~,
\eeq
where the second term is the Schwarzian derivative
\beq
\{x^\pm,y^\pm\} = \frac{{x^\pm}'''}{{x^\pm}'} - \frac32\left(\frac{{x^\pm}''}{{x^\pm}'}\right)^2 ~,
\eeq
and the prime denotes a derivative with respect to $y^\pm$.

To give an example of this procedure, we explicitly consider two different vacuum states that will be of importance later on. First, let us define two different sets of null coordinates. To define the so-called static vacuum, the state natural for a static observer, we define null coordinates as
\beq
\sigma^\pm = t \pm r_* ~,
\eeq
where $r_*$ is a tortoise coordinate defined as
\beq
r_* = \int_0^r\rmd r'\left(1-r'^2/\ell^2\right)^{-1} = \ell\, \text{arctanh}(r/\ell)~.
\eeq
The metric in these coordinates is
\beq
\rmd s^2 = -\text{sech}^2\left(\frac{\sigma^+-\sigma^-}{2\ell}\right)\rmd\sigma^+\rmd\sigma^- ~,
\eeq
which only covers the static patch.

A different set of coordinates that can be used to define the Bunch-Davies or Hartle-Hawking vacuum are the affine parameters for the past and future horizon. With respect to the static coordinates defined previously these are given by
\beq \label{eq:KruskalMetric}
\bal
x^+ &= +\ell e^{+\sigma^+/\ell} = +\ell e^{+t/\ell}\sqrt{\frac{\ell-r}{\ell+r}} ~,\\
x^- &= -\ell e^{-\sigma^-/\ell} = -\ell e^{-t/\ell}\sqrt{\frac{\ell-r}{\ell+r}} ~.
\eal
\eeq
In this coordinate system the metric reads
\beq
\rmd s^2 = -\frac{4\ell^4}{(\ell^2-x^+x^-)^2}\rmd x^+\rmd x^- ~.
\eeq
These different coordinates and the part of the global Penrose diagram of two-dimensional de Sitter space they cover are indicated in Figure \ref{fig:PenroseRectangle}.
The static vacuum can now be defined by expanding the field operator in modes that are positive frequency with respect to $\sigma^\pm$ and per definition the vacuum expectation value of the diagonal components of the stress tensor vanish.
\beq
\braket{T_{\pm\pm}(\sigma^\pm)}_{\rm S} = 0 ~.
\eeq
Clearly, the static observer does not see any radiation in this state. Using \eqref{eq:stresstrafo}, we find that when expressed in the $x^\pm$ coordinates the static vacuum is singular.
\beq
\bal
\braket{T_{\pm\pm}(x^\pm)}_{\rm S} &= -\frac{c}{48\pi {(x^\pm)}^2} ~.
\eal
\eeq
The singularities at the future and past horizon are similar to the singularities in the Rindler vacuum in flat space or the Boulware vacuum for a black hole\footnote{In fact, because in two dimensions every metric is conformally flat, vacuum states in different spacetimes are simply related by a conformal transformation \cite{Candelas:1978gf}. The Rindler vacuum is conformally related to the static vacuum and the Minkowski vacuum is conformally related to the Bunch-Davies vacuum.} and they signal that formally an infinite amount of energy is required to prepare this state.

Following the same procedure as before, the Bunch-Davies state is defined by expanding the field operator in modes that are positive frequency with respect to the $x^\pm$ coordinates and the diagonal components of the vacuum expectation value of the stress tensor are
\beq
\braket{T_{\pm\pm}(x^\pm)}_{\rm BD} = 0 ~.
\eeq
In contrast with the static vacuum, in these coordinates this state is regular at the past and future horizon. Also, transforming this expression in static coordinates we find
\beq
\braket{T_{\pm\pm}(\sigma^\pm)}_{\rm BD} = \frac{c}{48\pi \ell^2} ~.
\eeq
The non-zero value shows that a static observer measures a flux of both left and right-moving radiation at an inverse temperature $\beta=2\pi\ell$. Because there is an equal amount of left-moving and right-moving radiation, the Bunch-Davies state is a thermal equilibrium state.

Having discussed these two vacua, it is now straightforward to define a non-equilibrium vacuum state where the left and right-moving flux do not have the same temperatures. Such a state breaks the thermal equilibrium and is appropriate to describe the situation where radiation emitted by the cosmological horizon is collected inside the static patch, but no or less radiation escapes it by means of an observer. This leads to a pileup of positive energy inside the static patch that a static observer can access and shrinks the cosmological horizon \cite{Aalsma:2019rpt}. The general expression for the stress tensor expressed in $\sigma^\pm$ coordinates is
\beq
\braket{T_{\pm\pm}(\sigma_\pm)} = \frac{\pi c}{12\beta^{2}_\pm} ~,
\eeq
where $\beta_\pm$ are the different inverse temperatures of radiation at a constant $\sigma^\pm$ slice at the center of the static patch.

We are now interested in the situation where radiation is emitted from the past cosmological horizon into the static patch and less (or no) radiation is leaving it. Focusing on the ``right'' static patch in  Figure \ref{fig:PenroseRectangle} this is achieved by setting $\beta_-=2\pi\ell$ and taking $0\leq \beta_+^{-1} < 2\pi\ell$. This leads to a state that is singular at the past cosmological horizon $x^+=0$. As mentioned before, this singularity is not unexpected and simply reflects the infinite total flux present in this state, similar to the singularity in the Unruh vacuum of a black hole. In a more sophisticated treatment, which we will not pursue here, one could start out in the Bunch-Davies state and slowly allow an observer to collect radiation such that the total flux remains finite. 

Before we discuss backreaction, we stress some differences between the entanglement structure of the Bunch-Davies and the static vacuum. The (pure) Bunch-Davies state can be written as a maximally entangled (thermofield double) state between the two static patches surrounding the North and South pole. The density matrix is given by \cite{Goheer:2002vf}
\beq
\rho = \frac1Z\sum_i e^{-\beta \omega_i} \ket{\omega_i}_{\rm L}\ket{\omega_i}_{\rm R}\bra{\omega_i}_{\rm L}\bra{\omega_i}_{\rm R} ~,
\eeq
where $Z$ is a normalization factor, $\beta$ the inverse de Sitter temperature, and $L,R$ denote the different static patches. Tracing out one of the static patches we obtain a state with a thermal reduced density matrix
\beq
\rho_{\rm R} = \text{tr}_{\rm L}\,\rho = \frac1Z \sum_i e^{-\beta \omega_i}\ket{\omega_i}_R\bra{\omega_i}_R ~,
\eeq
at an inverse temperature $\beta = 2\pi\ell$. This state is maximally entangled with the static patch we traced out and because of the precise entanglement structure regular everywhere.\footnote{Per definition of the reduced density matrix, the vacuum expectation value of any local operator evaluated in the static patch can be computed as $\braket{{\cal O}}=\text{tr}_{\rm L}({\cal O}\rho_{\rm R})$. Contrary to some claims in the literature \cite{Markkanen:2017abw,Blumenhagen:2020doa}, this does not break the isometries of the state under consideration. This will be explained more in detail in \cite{Aalsma:2021xxx}.} In contrast, the static vacuum is obtained by taking $\beta \to \infty$ which results in a (pure) state that factorizes. As a result, this state exhibits no entanglement between the poles leading to singularities at both the past and future cosmological horizon.

This relates to the non-equilibrium state we are interested in as follows. While matter moving in the $x^+$ direction (right-moving radiation) is in the Bunch-Davies state, matter moving in the $x^-$ direction (left-moving radiation) interpolates between the static and Bunch-Davies vacuum depending on the temperature we specify. The right-moving radiation is purified by including the region beyond the future cosmological horizon, but the left-moving sector is already pure in one static patch when $\beta_+\to\infty$. If $\beta_+^{-1}>0$ we need to include the region beyond the past cosmological horizon to obtain a pure state. These observations will be important later on when we compute the fine-grained entropy.

\subsection{Backreacted solution}
Now that we have constructed a set of non-equilibrium states with different left and right-moving temperatures we are interested in studying their backreaction. Before doing so, we first briefly treat the more conventional case of the Bunch-Davies vacuum. Taking $\beta_\pm = 2\pi\ell$ we find that the equations of motion \eqref{eq:EOM} are solved by (for both the half and full reduction)
\beq
\Phi(t,r) = \frac1{24}\left(c+\Phi_s\frac{r}{\ell} \right)~.
\eeq
We see that the solution is still static and adding conformal matter only shifts the vacuum solution by $c/24$. As explained before, we work in a regime where $1\ll c \ll \Phi_s\ll \Phi_0 $. In a non-equilibrium state with $\beta_-=2\pi\ell$ the diagonal components of the stress tensor expressed in $x^\pm$ coordinates are given by
\beq \label{eq:UnruhStress}
\bal
\braket{T_{++}(x^+)} &= -\frac{c}{48\pi {(x^+)}^2}\left(1-t_+^2\right) ~, \\
\braket{T_{--}(x^-)} &= 0 ~. \\
\eal
\eeq
Here $t_\pm=2\pi\ell/\beta_\pm$ is the rescaled temperature. In this state, the equations of motion are solved by
\beq \label{eq:dilatonsolution}
\Phi(x^+,x^-) = \frac{c}{48}\left[1+\frac{2\Phi_s}{c\ell}r + t_+^2 -\left(1-t_+^2\right) \frac{r}{\ell}\log(x^+/\ell)\right] ~,
\eeq
where we used
\beq
\frac r{\ell} = \frac{\ell^2+x^+x^-}{\ell^2-x^+x^-} ~.
\eeq
We recognize the term proportional to $\Phi_s$ in \eqref{eq:dilatonsolution} as the vacuum solution. The logarithmic term causes the solution to no longer be static, which is a direct consequence of breaking the thermal equilibrium of the Bunch-Davies state. An interesting aspect of this solution is that the dilaton not only diverges at $x^+x^-=\ell^2$ as in the Bunch-Davies state, but also at $x^+\to0$ and $x^+\to\infty$. From a four-dimensional perspective, this can be intuitively understood as follows. Increasing the positive energy inside the static patch causes the black hole horizon to grow and the cosmological horizon to shrink. Hence, in the far past the size of the $S^2$ measured on the cosmological horizon $x^+=0$ was large which is reflected by the diverging dilaton. In contrast, in the far future it is the $S^2$ on the black hole horizon that becomes large and the dilaton therefore diverges at $x^+\to\infty$. The time $t=0$ is now special in the sense that the size of the dilaton on both horizons is approximately equal to the size it has in the Bunch-Davies state. Because the dilaton is inversely proportional to the gravitational coupling we find that gravity is weak on both horizons. We will leverage this feature later on to compute the fine-grained entropy.

\begin{figure}[h]
\centering
\includegraphics[scale=1]{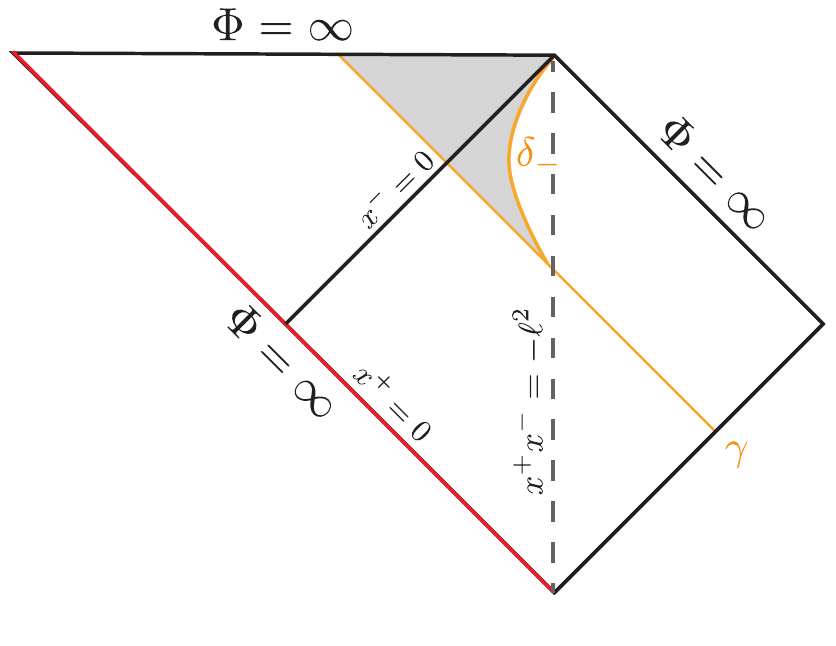}
\caption{Part of the Penrose diagram covered by $x^\pm$ coordinates in the non-equilibrium state \eqref{eq:UnruhStress} with $t_+=0$. The red line indicates the stress tensor singularity at $x^+=0$. The region shaded gray is a trapped region.}
\label{fig:PenroseUnruh}
\end{figure}

After including backreaction, the location of the horizons can be deduced from the behaviour of the dilaton. In higher dimensions, we can define two expansion scalars $\theta_\pm$ whose values characterize the behaviour of null congruences (see e.g. \cite{Poisson:2009pwt}). In JT gravity, the analogous quantities are the null derivatives of the dilaton: $\partial_\pm\Phi$. Using the explicit solution \eqref{eq:dilatonsolution} we can classify the different regions in which the expansion scalars have distinct behaviour in terms of the following three curves.
\beq
\bal
\gamma: \quad x^+ &= \ell \exp\left(\frac{2\Phi_s}{c(1-t_+^2)}\right) ~, \\
\delta_\pm:\quad x^-&= \frac{\ell^2}{x^+(1-t_+^2)}\left[(1-t_+^2)\log(x^+/\ell)-\frac{2\Phi_s}{c} \right. \\
&\left.\pm \sqrt{(1-t_+^2)^2 +c^{-2}\left(2\Phi_s-c(1-t_{+}^2)\log(x^+/\ell)\right)^2 } \right] ~.
\eal
\eeq
The apparent horizon (the outermost surface with $\partial_+\Phi=0$ and $\partial_-\Phi<0$) is given by the portion of the curve $\delta_-$ with $x^+>x^+_{\gamma}$. The location where  $\partial_+\Phi=\partial_-\Phi=0$ is given by $(x^+,x^-)=(x^+_{\gamma},-\ell^2/x^+_{\gamma})$ which is the pole. As measured by an observer at the pole, a trapped region ($\partial_\pm \Phi <0$) develops after a static time
\beq \label{eq:trapscale}
t_{\rm trap} = \frac{2\Phi_s\ell}{c(1-t_+^2)} ~.
\eeq
The backreacted geometry is showed in Figure  \ref{fig:PenroseUnruh}. As required, for $t_{+}=1$ the trapped timescale blows up since the corresponding Bunch-Davies state is static.

\section{Thermodynamics and fine-grained entropy} \label{sec:entropy}
Having discussed the backreacted geometry in a non-equilibrium state where (some amount of) radiation is collected inside the static patch, we now come back to our original question. What are the implications of the finiteness of the de Sitter entropy for a static observer? 
Can an observer decode information by performing measurements on the radiation collected inside the static patch? If so, the fine-grained entropy of the radiation should follow a Page curve, initially increasing when the size of the subsystem grows, but decreasing after the Page time.

Using the island formula, we compute the fine-grained entropy of radiation collected in the static patch in the two different models discussed in the previous section: the full and half reduction. For the island formula to be valid, we have to apply it to a region where gravity is weak which corresponds to large $\Phi$. Before computing the fine-grained entropy, we first set expectations by using thermodynamics to estimate the behaviour of the entropy and the Page time. After that, we reproduce those features using the island formula. 
We find that in the full reduction, an island forms at a time $t_{\rm Island}\simeq \Phi_s\ell/c$ which is parametrically smaller than the Page time $t_{\rm Page} \simeq \Phi_0\ell/c$ in a regime of semi-classical control (precise expressions will be given later on). This results in a Page curve for the Hawking radiation and allows for information recovery. In the half reduction on the other hand, $\Phi_0=0$ and there is no hierarchy between the time the island forms and the time when the horizon entropy would have been depleted. As a result information recovery does not seem possible in this setup.

\subsection{Thermodynamics}
Let us consider a conformal matter sector that consists of a gas of $c$ different bosonic species. The free energy of such a theory on a line of length $L$ can be easily computed using standard methods of statistical physics, see e.g. \cite{TongLectures}, and is given by
\beq
F =  -\frac{c\pi}{12\beta^2}L ~.
\eeq
From the free energy all other thermodynamic quantities can be computed, such as the energy density ${\cal E}$ and entropy density ${\cal S}_r$.
\beq
\bal
{\cal E} &= L^{-1}\partial_\beta(\beta F)=\frac{c\pi}{12\beta^2} ~, \\
{\cal S}_r &=L^{-1}\beta^2\partial_\beta F = \frac{c\pi}{6\beta} ~.
\eal
\eeq
We are interested in the energy and entropy that is collected inside of the static patch, which corresponds to the difference in the right-moving and left-moving contribution. Thus, in the non-equilibrium state with $t_-=1$ the net energy and entropy density are given by
\beq
\bal
{\cal E} &= \frac{c}{48\pi\ell^2}(1-t_+^2) ~, \\
{\cal S}_r &= \frac{c}{12\ell}(1-t_+) ~.
\eal
\eeq
As expected, in the Bunch-Davies state ($t_+=1$) no net entropy nor energy is collected in the static patch which is the property that makes it an equilibrium state. This contrasts the analysis of \cite{Sybesma:2020fxg} where the entanglement of the entire static patch with its complement was studied, rather than an observer who collects large amounts of radiation.

Integrating the entropy density we find that away from the thermal equilibrium the entropy grows linearly as a function of static time
\beq \label{eq:radiationentropy}
S_r = \frac{c}{12\ell}(1-t_+)t ~.
\eeq
By setting the integration constant to zero, we implicitly impose that $t=0$ is the time when an observer starts collecting radiation in the static patch. Of course, in the way we have set up the state \eqref{eq:UnruhStress} there is also a net flux of energy before this time. However, $t=0$ is special in the sense that at that time the entropy of the cosmological horizon is approximately equal to the entropy in the Bunch-Davies state.

In addition to the increasing radiation entropy, the horizon carries a gravitational entropy proportional to the area which shrinks due to the backreaction of positive energy piling up in the static patch. At $t=0$, the gravitational entropy measured by the cosmological horizon is given by $2\Phi_{H}(t=0)$. Expressed in terms of coordinates natural for an observer at the pole, the horizon entropy is
\beq \label{eq:HorizonEntropyThermo}
S_H = 2\Phi_H(t) = 2\Phi_0+\frac c{24}\left(1+t_+^2+\frac{2\Phi_s}{c}\right) - \frac{c}{24\ell}(1-t_+^2)t~.
\eeq
Here we evaluated the dilaton on the future cosmological horizon and used $x^+=\ell e^{t/\ell}$. The time it takes to deplete the horizon entropy entirely is given by
\begin{equation}
    t_{\text{end}}
    =
    \frac{48\ell}{c(1-t_{+}^{2})}
    \Phi_H(t=0)
    \,.
\end{equation}
Comparing the horizon entropy to the radiation entropy we find that, from a thermodynamic perspective, the Page time is given by
\beq\label{eq:page_time_thermo}
t_{\rm Page} = \frac{48\ell}{c(3-2t_+-t_+^2)}\Phi_H(t=0) 
=
    \frac{
        1-t_{+}^{2}
    }{
        3-2t_{+}-t^{2}_{+}
    }
    t_{\text{end}}
~.
\eeq
Of course, this expression is only valid away from $t_+=1$ since there is no net radiation collected in the Bunch-Davies state.

\subsection{Fine-grained entropy}
We will now compute the fine-grained entropy of radiation collected in a region $R$ where gravity is weak. The fine-grained entropy of this region according to the island formula is given by
\beq
S(R) = \text{min ext}_I\left[2\Phi_H(\partial I) + S_{\rm vN}(R\cup I)\right] ~.
\eeq
Here we are instructed to extremize the location of the endpoint of the island $\partial I$ and minimize the entropy over all possible islands. Because we want our global state to be pure, we only consider the non-equilibrium state with $t_+=0$. As we discussed in the previous section, doing so allows us to ignore the region beyond the past cosmological horizon to obtain a pure state. This is necessary because the dilaton diverges at the past horizon, obstructing a continuation beyond it within two-dimensional JT gravity. We will now assess the possible contribution of islands in the two models we discussed: the full and half reduction.

\subsubsection*{Full reduction}
\begin{figure}[h]
\centering
\includegraphics[scale=1]{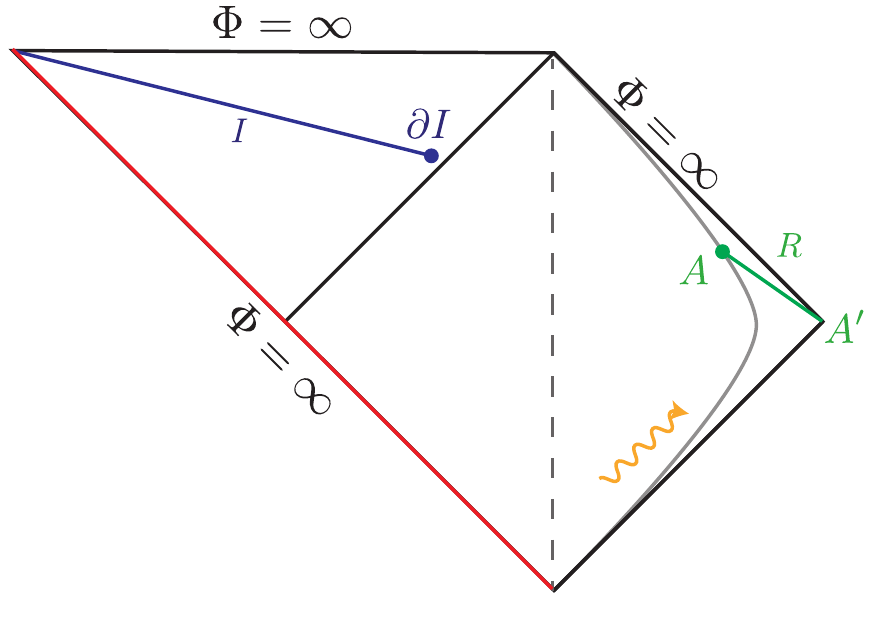}
\caption{Using the formula for the fine-grained entropy, we compute the entropy $S(R)$ of a weakly-gravitating region $R=[A,A']$ which contains radiation emanating from the past horizon. Because the radiation is entangled with the region beyond the future horizon, we allow for a possible island $I$ that contributes to $S(R)$.}
\label{fig:PenroseIsland}
\end{figure}

In the full reduction we found that $\Phi\to\infty$ in the limit $x^+\to\infty$. We can therefore define an interval $R$ in a weakly gravitating region by considering an anchor curve parametrized by
\beq
x^\pm = \pm \hat r e^{\pm t/\ell} ~.
\eeq
We take $\hat r/\ell$ sufficiently large such that the dilaton obeys $\Phi\gg 1$ in $R$ and gravity is weak. The endpoints $A$ and $A'$ of the interval $R$ are now defined to lie at the anchor curve and $x_{A'}^+/\ell=-x_{A'}^-/\ell\to\infty$ respectively. This way, we have defined a region where the fine-grained entropy emitted by the cosmological horizon can be computed in a controlled manner, see Figure \ref{fig:PenroseIsland}.

To compute the von Neumann entropy $S_{\rm vN}(R)$ we make us of standard CFT results. Writing the metric in conformal gauge
\beq
\rmd s^2 = -\Omega^{-2}\rmd y^+\rmd y^- ~,
\eeq
the von Neumann entropy of a CFT on an interval $[y_1,y_2]$ in its vacuum state is given by \cite{Fiola:1994ir,Almheiri:2019hni}
\beq \label{eq:vNgeneral}
S_{\rm vN}(y_1,y_2) = \frac c{12}\log\left[\frac{(y_1^+-y_2^+)^2}{\epsilon^2\Omega(y_1)\Omega(y_2)}\right]+\frac c{12}\log\left[\frac{(y_1^--y_2^-)^2}{\epsilon^2\Omega(y_1)\Omega(y_2)}\right] ~,
\eeq
which is a sum of the entropy of left-movers and right-movers. Here $\epsilon$ is a regulator that will not play an important role since we'll be comparing contributions from different saddles of the generalized entropy. When $t_+=0$ the $y^\pm(x^\pm)$ coordinates that define the vacuum are related to $x^{\pm}$ as
\beq
y^-(x^-) = x^- \quad \text{and} \quad y^+(x^+) = \ell \log(x^+/\ell)~.
\eeq
In the $y^\pm$ coordinates the metric becomes
\beq
\rmd s^2 = -\frac{4\ell^3x^+}{(\ell^2-x^+x^-)^2}\rmd y^+\rmd y^- ~,
\eeq
from which we can read off the conformal factor
\beq
\Omega^{-2} = \frac{4\ell^3 x^+}{(\ell^2-x^+x^-)^2} ~.
\eeq
Then, in this state the von Neumann entropy is given by
\beq \label{eq:vNRnoIsland}
\bal
S_{\rm vN}(R) &= \frac c{12}\log\left[\frac{4\ell^3(x^+_{A}x^+_{A'})^{1/2}(x_{A}^--x_{A'}^-)^2}{\epsilon^2(\ell^2-x_A^+x_A^-)(\ell^2-x_{A'}^+x_{A'}^-)}\right] \\ &+\frac c{12}\log\left[\frac{4\ell^5(x^+_{A}x^+_{A'})^{1/2}\log(x_{A'}^+/x_{A}^+)^2}{\epsilon^2(\ell^2-x_A^+x_A^-)(\ell^2-x_{A'}^+x_{A'}^-)}\right] ~.
\eal
\eeq
We now express the von Neumann entropy in static coordinates using
\beq
x^\pm = \pm\ell e^{\pm t/\ell}\sqrt{\frac{\ell-r}{\ell+r}} ~,
\eeq
and take a derivative with respect to $t$ in order to remove unimportant additional constants. In the limit $x_{A'}^+=-x_{A'}^-\to\infty$ we then find
\beq \label{eq:vNRsimple}
S_{\rm vN}(R) =\frac c{12\ell}t + \dots ~,
\eeq
where the dots denote unimportant time-independent contributions. This expression agrees with the thermodynamic entropy (setting $t_+=0$) and grows without bound.

Of course, according to the island formula we should allow for a possible island in which case the fine-grained entropy is given by
\beq \label{eq:IslandFormula}
S(R) = \text{min ext}_I\left[2\Phi_H(\partial I) + S_{\rm vN}(R\cup I) \right]~.
\eeq
To find the location of the endpoint $\partial I$ of the island we need to extremize $S(R)$. In a pure state, the von Neumann entropy of the region $R\cup I$ is given by the entropy on $[\partial I,A]$ which takes the form
\beq
\bal
S_{\rm vN}(R\cup I) &= \frac c{12}\log\left[\frac{4\ell^3(x^+_{\partial I}x^+_{A})^{1/2}(x_{\partial I}^--x_{A}^-)^2}{\epsilon^2(\ell^2-x_{\partial I}^+x_{\partial I}^-)(\ell^2-x_{A}^+x_{A}^-)}\right]
\\&+\frac c{12}\log\left[\frac{4\ell^5(x^+_{\partial I}x^+_{A})^{1/2}\log(x_A^+/x_{\partial I}^+)^2}{\epsilon^2(\ell^2-x_{\partial I}^+x_{\partial I}^-)(\ell^2-x_{A}^+x_{A}^-)}\right] ~.
\eal
\eeq
For a vanishing island we reproduce \eqref{eq:vNRsimple}. To find an analytical expression for the extremum $\partial_\pm S(R)=0$, we will assume that the second term in $S_{\rm vN}(R\cup I)$ is negligible. We expect this approximation to be reasonable because that term captures the contribution of modes moving in the $x^-$ direction which are in the static vacuum that exhibits no superhorizon entanglement. We confirm numerically in  Appendix \ref{app:numerical} that corrections to this approximation are small. 

Extremizing the fine-grained entropy this way we find the following solution for a non-trivial island.
\beq \label{eq:IslandLoc}
(x_{\partial I}^+,x_{\partial I}^-) = \left(-\frac{2\ell^2}{x_A^-W_0\left[f(x_A^-)\right]},0\right) ~.
\eeq
Here $W_0[x]$ is the principal branch of the Lambert $W$ function and
\beq
f(x^-)=-\frac{2\ell e^{-1-2\Phi_s/c}}{x^-} ~.
\eeq
Expressing $x_A^-$ at the pole as $x_A^-=-\ell e^{-t_A/\ell}$ we find that for $t_A/\ell \gg 1$
\beq
\log(x_{\partial I}^+/\ell) = t_A/\ell + \dots ~,
\eeq
where the dots denote logarithmic corrections. In the approximation we used, we find $x^{-}_{\partial I}=0$. We have verified numerically that without making this approximation the island sits close to but slightly behind the cosmological horizon, i.e. $x^{-}_{\partial I}>0$.

\begin{figure}[ht]
\centering
\includegraphics[scale=.6]{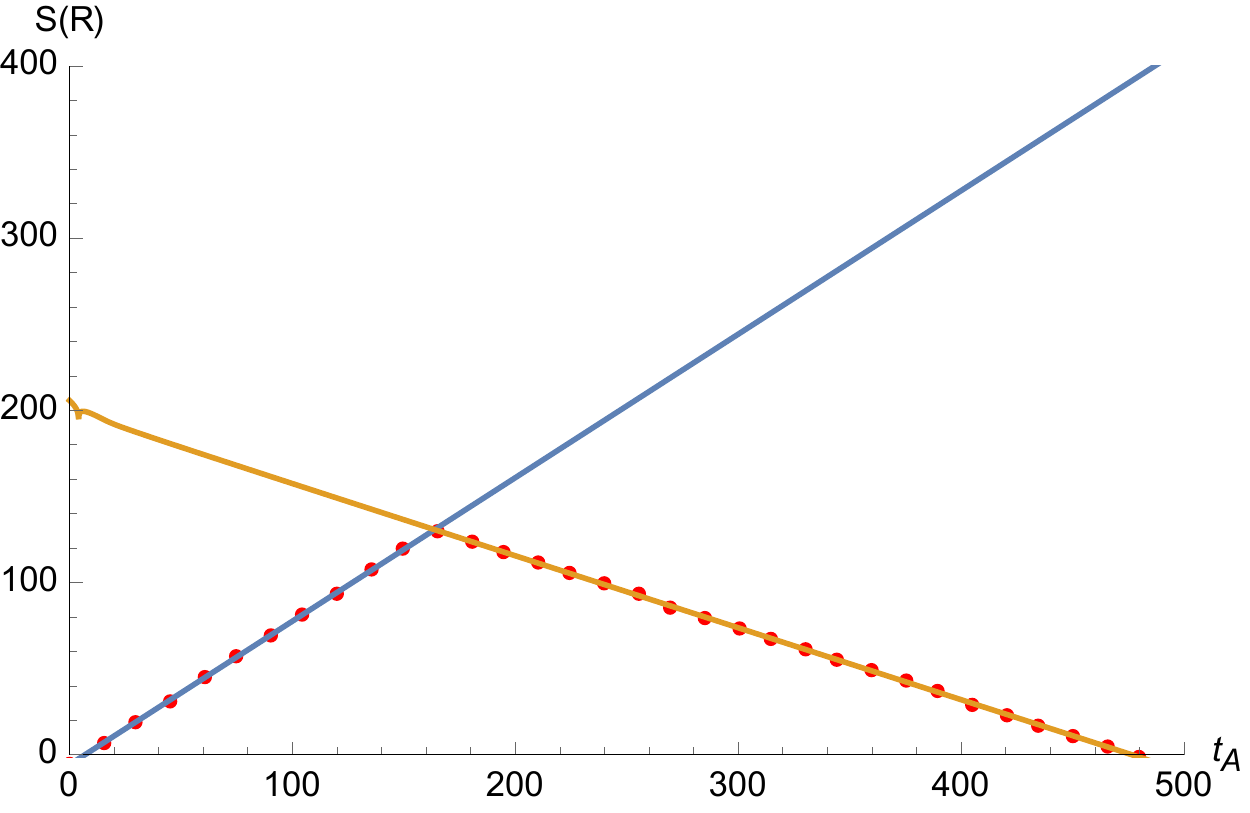}
\caption{The fine-grained entropy $S(R)$ in the non-equilibrium state in the full reduction. The solid blue line corresponds to the entropy evaluated on the trivial island \eqref{eq:vNRsimple} and the decreasing orange line to the entropy evaluated on the non-trivial island \eqref{eq:IslandLoc}. The fine-grained entropy is given by the minimum of both contributions indicated by the red dotted line. To produce the figure we took $\ell=1, \Phi_s=50, \Phi_0=100, c=10, \hat r/\ell = 10^3$.}
\label{fig:PageCurve}
\end{figure}

Comparing the trivial and non-trivial island we find that initially the trivial island gives the lowest entropy, but as the cosmological horizon decreases the non-trivial island eventually becomes dominant. We show the behaviour of $S(R)$, which follows a Page curve, as a function of static time in Figure \ref{fig:PageCurve}.
The decreasing entropy of the non-trivial island closely follows the decrease of the thermodynamics horizon entropy computed in \eqref{eq:HorizonEntropyThermo} and the increasing entropy of the trivial island follows \eqref{eq:vNRsimple}. As such, the Page time is easily computed to be
\beq
t_{\rm Page} = \frac{16\ell}{c}\Phi_H(t=0) ~.
\eeq
After the Page time, information behind the future cosmological horizon can be decoded by performing measurements on the subsystem containing the collected radiation \cite{Hayden:2007cs}. However, as we show in Section \ref{sec:fate} the collected radiation will eventually backreact strongly leading to the formation of a singularity.

\subsubsection*{Half reduction}
\begin{figure}[h]
\centering
\includegraphics[scale=1]{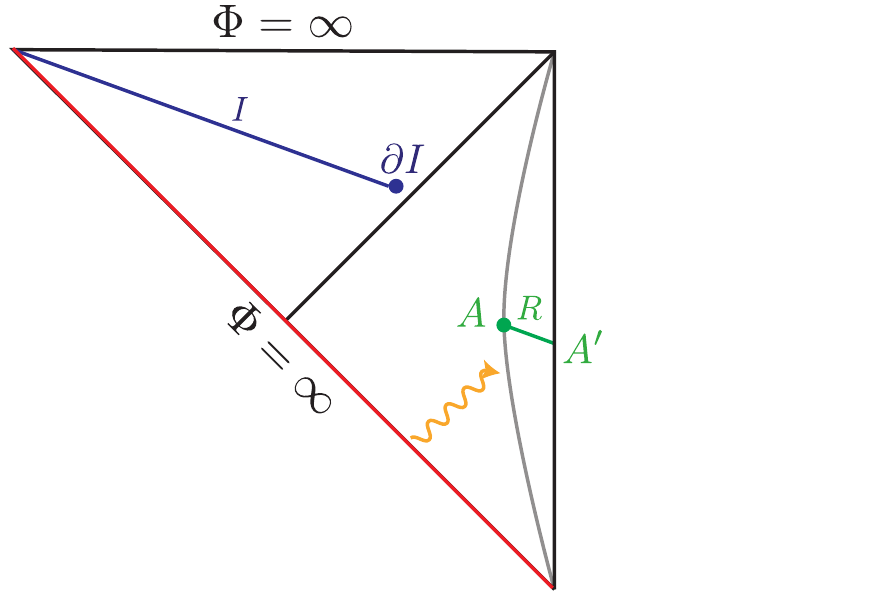}
\caption{Using the formula for the fine-grained entropy, we compute the entropy $S(R)$ of the region $R=[A,A']$ defined by the gray anchor curve which contains radiation emanating from the past horizon. Gravity is weak in the region $R$. Because the radiation is entangled with the region beyond the future horizon, we allow for the possibility that an island $I$ contributes to $S(R)$.}
\label{fig:PenroseIslandHalf}
\end{figure}

We now consider the half reduction and define a region $R$ by picking an anchor curve that lies very close to the pole of de Sitter, see Figure \ref{fig:PenroseIslandHalf}.
In the state with $t_+=0$, at the pole the dilaton takes the value
\beq
\Phi = \frac{c}{48} ~,
\eeq
so when $c\gg 1$ gravitational effects are small. In this state, all radiation emanating from the past horizon is collected in the region around the pole such that no left-moving radiation is present. The computation of the fine-grained entropy now closely parallels the computation in the full reduction. The von Neumann entropy of the region $R$ is given by \eqref{eq:vNRnoIsland} with $x^+_{A'}=-x^-_{A'}=\ell$ and the location of the island coincides with that of the full reduction. Note however that in the limit $x^-_A \to-\infty$ the location of the island plateaus to a fixed location.
\beq
\lim_{x_A^-\to-\infty}x_{\partial I}^+ = \ell e^{1+2\Phi_s/c} ~.
\eeq
This implies that with respect to a static observer at the pole, the island is only spacelike separated when $t \gtrsim 2\Phi_s\ell/c$ and the island can only contribute after this time. Because $\Phi_0=0$ in the half reduction, there is no parametric separation between the time the island forms and the Page time. The consequence of this is that when the island forms, $\Phi\simeq 0$ at the location of the island signalling strong coupling. For this reason, it does not seem possible in this model to recover information in a controlled manner by absorbing Hawking radiation. 

\subsubsection*{Scrambling time}
Before we close this section and discuss the fate of an observer that collects and decodes radiation, we briefly mention how long it takes to recover a small amount of information that is thrown through the cosmological horizon after the Page time in the full reduction. For black holes, this is the so-called scrambling time \cite{Sekino:2008he,Hayden:2007cs,Shenker:2013pqa} which is given by $t_*\simeq \frac{\beta}{2\pi}\log(S_{BH})$ saturating the chaos bound \cite{Maldacena:2015waa}. $S_{BH}$ is the black hole entropy and $\beta$ its inverse temperature.

To determine the recovery time in the present context, we compute the time difference between a future-directed light ray emitted from the pole that precisely hits the island when it crosses the horizon and a past-directed lightray emitted from the point $A$ in $R$, see Figure \ref{fig:PenroseIsland}. This difference gives the amount of time it takes for the information to be transferred from the entanglement wedge of the island to the entanglement wedge of $R$ and therefore determines the recovery time. Using the expression for the location of the island we find that around the Page time this time is given by
\beq
t_* = \ell \log(S_H) ~,
\eeq
up to subleading corections. Here $S_H$ is the horizon entropy given by \eqref{eq:HorizonEntropyThermo}. As for black holes, this expression coincides with the scrambling time in de Sitter space which has been studied in \cite{Susskind:2011ap,Aalsma:2020aib,Blommaert:2020tht,Bhattacharyya:2020kgu,Haque:2020pmp,Bedroya:2020rmd,Lehners:2020pem,Sybesma:2020fxg}.

\section{Fate of the observer} \label{sec:fate}
In the previous section, we showed that the fine-grained entropy of radiation collected in a static patch follows a Page curve which is consistent with the idea that a static observer in de Sitter space should be able to recover information that has fallen through their cosmological horizon. However, because the static patch has finite volume beackreaction eventually has to become large and it is not guaranteed that an observer will survive this experiment.

\subsection{Quantum singularity theorem}
In the state with $t_+=0$ a non-trivial island appears after a time $t\simeq 2\Phi_s\ell /c$, as measured from the pole, showing entanglement between the collected radiation and the region beyond the future cosmological horizon. Interestingly, this is of the same order as the timescale $t_{\rm trap}$ (see \eqref{eq:trapscale}) after which a trapped region develops. This might suggest the formation of a singularity, but Penrose's classical singularity theorem \cite{Hawking:1973uf} does not apply here because the stress tensor \eqref{eq:UnruhStress} violates the Null Energy Condition for $t_+^2<1$.

Instead, we can use a quantum singularity theorem proposed and proven in \cite{C:2013uza} which assumes the Generalized Second Law (GSL).\footnote{There exist other quantum singularity theorems that don't assume the GSL \cite{Freivogel:2020hiz}.} Given a Cauchy surface $\Sigma$ in a semi-classical region, we define a codimension-two surface $P$ that splits $\Sigma$ in an interior part $\Sigma_{\rm in}$ and an exterior part $\Sigma_{\rm out}$. The generalized entropy at $P$ is now defined as
\beq
S(P) = \frac{\text{Area}(P)}{4G_N} + S_{\rm vN}(\Sigma_{\rm out}) ~.
\eeq
The first term is a Bekenstein-Hawking like term and the second term is the von Neumann entropy of the region $\Sigma_{\rm out}$.\footnote{This of course resembles the island formula, which can be thought of as the generalized entropy for disconnected regions evaluated on a quantum extremal surface.}

The GSL now states that $S(P)$ evaluated on future causal horizons is non-decreasing with time. The generalized entropy is only required to increase on future causal horizons and, in analogy with classically trapped regions, there can be quantum trapped regions where both null derivatives of the generalized entropy are negative. When $\Sigma$ is a non-compact Cauchy surface in a globally hyperbolic spacetime, it follows \cite{C:2013uza} that when a quantum trapped region forms the spacetime is geodesically incomplete. Just like Penrose's classical singularity theorem this implies the formation of a singularity. We now use this result to determine the fate of the observer.

\begin{figure}[ht]
\centering
\includegraphics[scale=1]{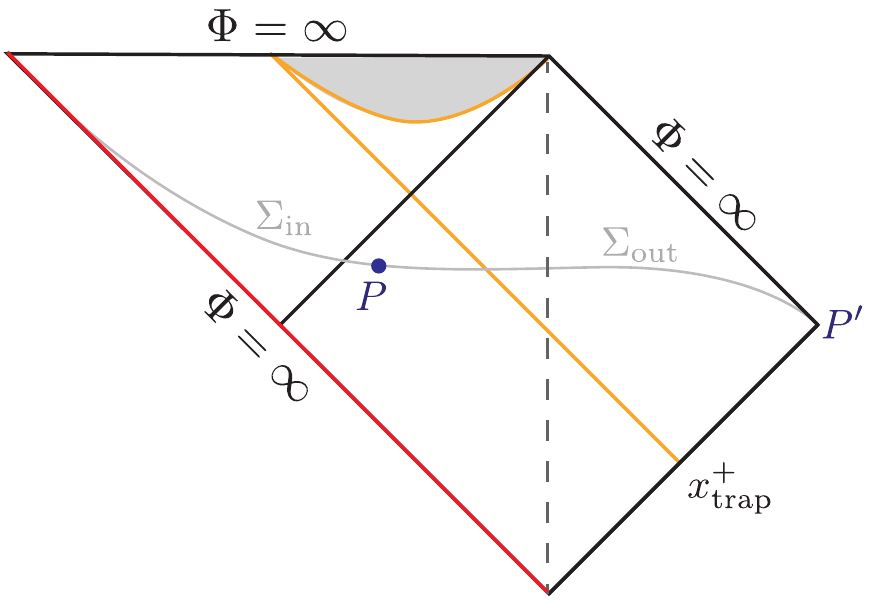}
\caption{To compute the generalized entropy $S(P)$ we define a Cauchy surface $\Sigma$ and consider a point $P$ that splits it into a region $\Sigma_{\rm in}$ and $\Sigma_{\rm out}$. Taking null derivatives of the generalized entropy we find a quantum trapped region (shaded gray) where the generalized entropy decreases in both null directions. The quantum trapped region develops much later than a scrambling time, but much earlier than the Page time: $t_{*}\ll t_{\rm trap}\ll t_{\rm Page}$.}
\label{fig:GenEnt}
\end{figure}

Using \eqref{eq:dilatonsolution} and \eqref{eq:vNgeneral} the generalized entropy in the $t_+=0$ state is given by
\beq
\bal
S(P) = 2\Phi_H(P) &+ \frac c{12}\log\left[\frac{4\ell^3(x^+_{P}x^+_{P'})^{1/2}(x_{P}^--x_{P'}^-)^2}{\epsilon^2(\ell^2-x_P^+x_P^-)(\ell^2-x_{P'}^+x_{P'}^-)}\right] \\
&+\frac c{12}\log\left[\frac{4\ell^5(x^+_{P}x^+_{P'})^{1/2}\log(x_{P'}^+/x_P^+)^2}{\epsilon^2(\ell^2-x_P^+x_P^-)(\ell^2-x_{P'}^+x_{P'}^-)}\right] ~.
\eal
\eeq
Taking derivatives with respect to $x^\pm$ we can express the different regions in terms of three curves. In the limit $x_{P'}^+ = -x_{P'}^-\to\infty$ we have
\beq
\bal
\zeta: \quad x^-_P&=\frac{\ell^2}{x^+_P}\left(1+\frac{\Phi_s}{c} -\frac12\log(x^+_P/\ell)\right)~,\\
\eta_\pm: \quad x^-_P&=\frac{\ell^2}{x^+_P}\left(\frac{2\Phi_s}{c}-\log(x_P^+/\ell) \pm \sqrt{1+\left(\frac{2\Phi_s}{c}-\log(x_P^+/\ell)\right)^2}\right) ~.
\eal
\eeq
We now find that a quantum trapped region forms after $x^+>x^+_{\rm trap}=\ell e^{2\Phi_s/c}$ in the region $x^->x^-_{\eta_+}$ and $x^->x^-_{\zeta}$, see Figure \ref{fig:GenEnt}.
The conclusion is that from the perspective of an observer at the pole a singularity will form after $t>t_{\rm trap}=2\Phi_s\ell/c$. Notice that the time when a quantum trapped region forms coincides with the formation of a classically trapped region. Thus, after this time it is guaranteed that the worldline of an observer will hit a singularity at future infinity.\footnote{If an observer would jump through the horizon at $x^+\to\infty$ in the Penrose diagram and thus enter a region beyond the regime of validity of our two-dimensional model, that observer would encounter the singularity of the black hole.} 

\subsection{No-cloning paradox}
The formation of a singularity is important to avoid the paradoxical situation where a bit of information is thrown through the horizon, after which an observer could decode it performing measurements on the Hawking radiation and then jump through the horizon to see the same bit of information twice. This would violate the quantum no-cloning theorem. In the present context, this is avoided the same way as for a black hole \cite{Hayden:2007cs,Susskind:1993mu}. To decode some information from the radiation takes a Page time, after which a new bit thrown through the horizon only takes a scrambling time to decode. In a regime of control $t_{\rm Page}\gg t_{\rm trap}$ any light ray that travels through the horizon after the Page time will terminate at a singularity. As a consequence, any observer that decodes a bit of information and tries to see its copy by jumping after it through the horizon will only have a short time to do so before it hits the singularity. Given the scrambling time $t_*=\ell \log(S_H)$ it turns out that this time is subplanckian, outside the regime of semi-classical control.

\section{Discussion} \label{sec:discussion}
In this paper, we studied information recovery in the static patch of de Sitter space by employing the island formula in two different two-dimensional JT gravity models with a known higher dimensional pedigree. We put particular emphasis on the role of a static observer, who only has access to a finite region of spacetime. As we explained, in order for a static observer to decode any information by performing measurements on collected radiation it is necessary to take into account backreaction. Positive energy collected in the static patch backreacts on the cosmological horizon, breaking the thermal equilibrium of the Bunch-Davies state typically considered. Our analysis crucially relied on a quantum state, first studied in \cite{Aalsma:2019rpt}, that describes the situation in which Hawking radiation emanating from the cosmological horizon is collected, but no radiation is passing the future cosmological horizon.

Using JT gravity, we solved for the backreacted geometry in this non-equilibrium state and used thermodynamics to compare the entropy of the radiation with the gravitational entropy of the horizon, showing that the radiation entropy naively seems to grow without bound. However, by using the island formula we found that in the JT gravity model we referred to as the full reduction an island forms at a time parametrically smaller than the Page time. In the other JT model we studied (the half reduction), we found that no island contributes to the entropy in a regime of semi-classical control. This can be traced back to the absence of a large topological term in this model. Taking the island in the full reduction into account, the fine-grained entropy of the radiation follows a Page curve. Of course, backreaction should eventually become large in our state since there is a continuous flux of energy being radiated in the finite static patch. We found that at the time the island appears, a quantum trapped region forms outside of the event horizon signalling the formation of a singularity. A static observer is guaranteed to hit this singularity, paying a large price for their curiosity. We argued that the formation of a singularity is required in order to avoid a quantum no-cloning paradox. Although the singularity forms outside of the event horizon, it seems appropriate to interpret the large backreaction caused by collecting too much radiation as collapsing the static patch. In the region $0<r<\ell$ the value of the dilaton indeed becomes large and negative for large $t$, signalling strong coupling.

Our results give an interpretation to the de Sitter entropy as the entropy associated to the horizon of a single static observer. By singling out a particular observer, the quantum state we used spontaneously breaks the isometry group of de Sitter space down to those that are preserved by a single static patch. In future work, it would be interesting to better understand what the implications are of our results for different observers related by an isometry. This is particularly relevant when we want to consider an inflationary setup where a meta-observer performs measurements on (a non-gravitational region glued to) ${\cal I}^+$ which (approximately) realizes all de Sitter isometries. Interestingly it has previously been suggested that, similar to the contribution of islands in black hole physics around the Page time, inflationary physics might receive corrections at long time scales \cite{ArkaniHamed:2007ky}. Although this is a tantalizing possibility, the precise role that islands play in inflationary setups has been a subject of debate \cite{Chen:2020tes,Hartman:2020khs}. Another possible avenue of future research would be to apply the approach developed in this work to a setup that includes an inflaton, see e.g. \cite{Manu:2020tty}.

Furthermore, our results support a notion of holography best articulated along the lines of complementarity which applied to our setup would suggests that gravitational physics in a static patch is holographically dual to a system living at the (stretched) cosmological horizon, which has a finite Hilbert space \cite{Susskind:1993if,Susskind:2021}. For this comparison to hold we find that it is essential that a singularity forms, associated to the trapped region which comes into play due to backreaction, in order to avoid cloning paradoxes and preserve unitarity.

In addition, it would be interesting to relate the single static observer we considered in this paper to two observers on opposite poles of de Sitter space inserting shockwaves. Doing so, brings opposite poles into causal contact \cite{Gao:2000ga,Aalsma:2020aib} allowing the exchange of information. An observer that acts with a shockwave on the maximally entangled Bunch-Davies state can exchange information between poles already in a scrambling time \cite{Aalsma:2021xxx} and not the longer Page time.

Finally, we would like to remark that in this paper we have purely focused on semi-classical gravity in de Sitter space leading to a controlled setup in which information recovery is possible. However, it has been suggested that in string theory the lifetime of an (approximate) de Sitter phase could be much shorter, see e.g. \cite{Garg:2018reu,Ooguri:2018wrx,Montero:2019ekk,Bedroya:2019snp,Montero:2020rpl,Blumenhagen:2020doa,Rudelius:2021oaz}. Our semi-classical approach does not seem to capture such effects and it would be interesting to understand if and how they could manifest themselves. We hope to come back to these questions in future work.

\section*{Acknowledgements}
We would like to thank Alek Bedroya, Gregory Loges, Jake McNamara, Tom Rudelius, Andy Svesko and Manus Visser for useful discussions. LA also thanks Alex Cole, Edward Morvan, Gary Shiu and Jan Pieter van der Schaar for collaboration on related work. LA is supported by the DOE under grant DE-SC0017647. WS is supported by the Icelandic Research Fund (IRF) via a Personal Postdoctoral Fellowship Grant (185371-051).

\appendix

\section{Numerical island} \label{app:numerical}
In this Appendix we numerically compute the fine-grained entropy in the full reduction to obtain location of the island without making any approximation. We compare this to the approximate solution \eqref{eq:IslandLoc}, analytically found by dropping the contribution of modes moving in the $x^-$ direction, and confirm that this approximation is good.

\begin{figure}[h!]
\centering
\includegraphics[scale=.6]{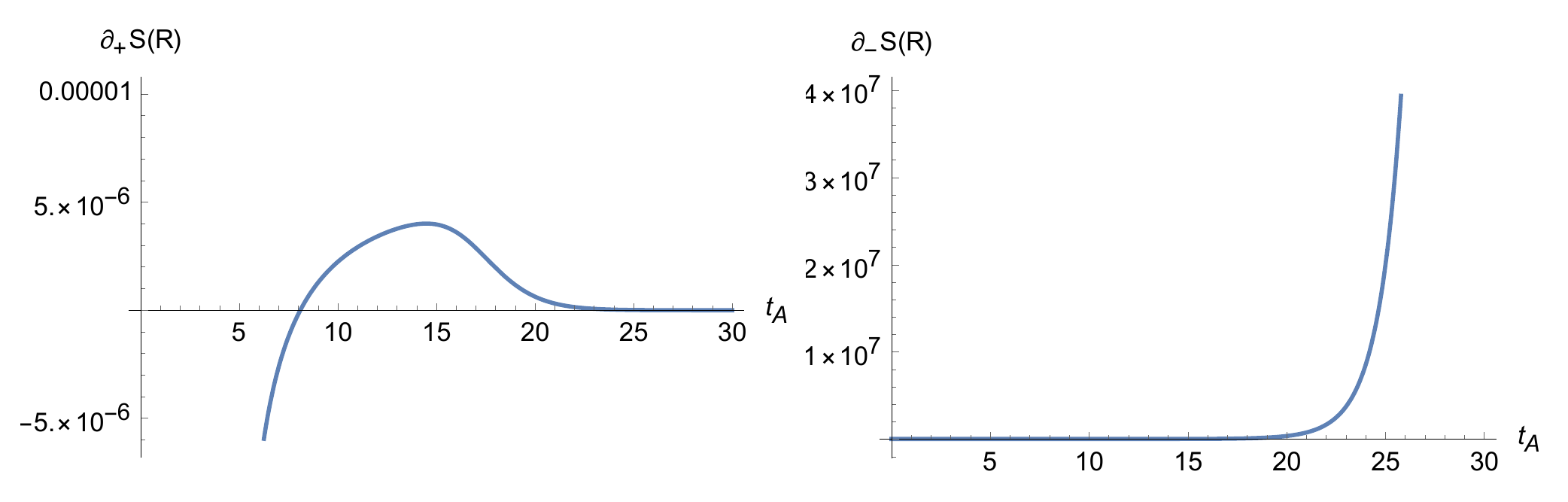}
\caption{$\partial_\pm S(R)$ evaluated on the approximate island. To produce the figures we took $\ell=1, \Phi_s=50, \Phi_0=100, c=10, \hat r/\ell = 10^3$.}
\label{fig:ExtremumError}
\end{figure}

\begin{figure}[h!]
\centering
\includegraphics[scale=.8]{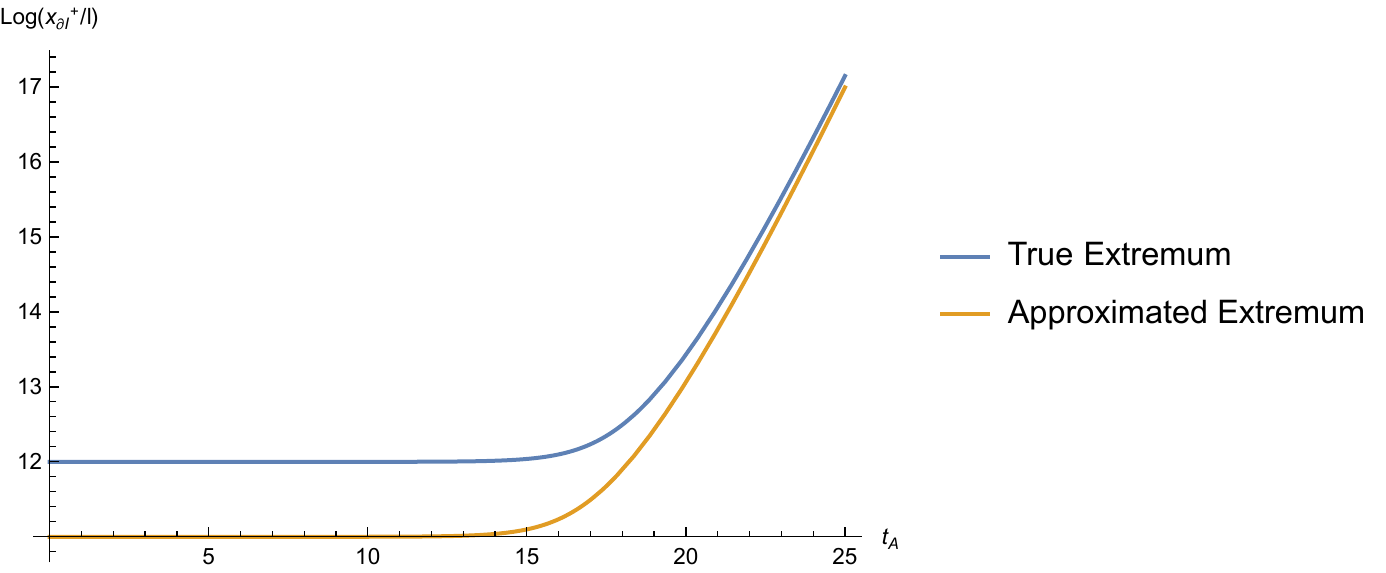}
\caption{Comparison of the approximate island and the ``true'' island found by solving $\partial_-S(R)$ at $x_{\partial I}^-=0$. To produce the figures we took $\ell=1, \Phi_s=50, \Phi_0=100, c=10, \hat r/\ell = 10^3$.}
\label{fig:CompareSols}
\end{figure}

To confirm that the approximate island given by
\beq
(x_{\partial I}^+,x_{\partial I}^-) = \left(-\frac{2\ell^2}{x_A^-W_0\left[f(x_A^-)\right]},0\right) ~,
\eeq
with
\beq
f(x^-)=-\frac{2\ell e^{-1-2\Phi_s/c}}{x^-} ~,
\eeq
is a good approximation to the true extremum, we first evaluate the full expression for $\partial_\pm S(R)$ (see \eqref{eq:IslandFormula}) on the approximate island. As we show in Figure \ref{fig:ExtremumError}, the error made in $S_+(R)$ quickly converges to zero after a time $t\simeq 2\Phi_s\ell/c$, so $x_{\partial I}^-=0$ is a good approximation. The error in $\partial_{-}S(R)$ on the other hand becomes large so we have to check in more detail that the true extremum is well approximated by \eqref{eq:IslandFormula}.
Solving $\partial_-S(R)$ evaluated on $x_{\partial I}^-=0$ for $x^+$ we find the ``true'' extremum and compare this with the approximate island. It turns out that again after a time $t\simeq 2\Phi_s\ell/c$ both solutions converge, see Figure \ref{fig:CompareSols}, confirming the validity of our approximation.

\bibliographystyle{utphys.bst}
\bibliography{refs}

\end{document}